\documentstyle[floats,multicol,aps,epsf,prl]{revtex}

\begin{document}
\draft
\twocolumn[\hsize\textwidth\columnwidth\hsize\csname @twocolumnfalse\endcsname
\def\btt#1{{\tt$\backslash$#1}}
\title{ Metastability and Glassy Behavior of a Driven Flux-Line Lattice.}
\author{W. Henderson  and E.Y. Andrei}
\address{ 
Department of Physics and Astronomy, Rutgers University,
Piscataway, NJ 08855}
\author{M.J. Higgins and S. Bhattacharya}
\address{NEC Research Institute , 4 Independence Way, Princeton, 
	 New Jersey 08540}
\maketitle
\begin {abstract}
{Strong metastability and history dependence are  observed in 
 DC and pulsed  transport studies  of 
flux-line lattices in  2H-$NbSe_{2}$, leading to the identification of
two distinct states of the lattice with different spatial ordering. 
The metastability is most pronounced upon crossing a transition line 
marked by a large jump in the critical current (the peak effect). Current
induced annealing of the metastable state towards the stable state is
observed with a strongly current dependent annealing time, which
diverges as a threshold current is approached from above.}
   
\end {abstract}
\pacs{PACS numbers: 74.60.Ge 74.60.Jg 74.60Ec }
]


In the absence of disorder the physics of a  magnetic flux-line lattice (FLL)
 is  governed by the interplay 
between thermal fluctuations, which favor  melting, 
 and  interactions, that   lead to ordering. The resulting 
phase diagram consists of a liquid and 
a solid phase with relatively simple dynamics. 
Quenched disorder causes the system  to develop additional 
phases and complex dynamic effects such as pinning and 
  irreversibility in the magnetic and electric responses. The role
played by disorder and pinning in the physics of FLL 
in equilibrium has 
 recently become an area of intense study \cite{blatter,majer}.
A related but distinct topic of  current interest 
concerns the role of motion on the spatial ordering of the FLL, the
resulting dynamical transitions or crossovers that may occur, and the
relation they bear to the disorder free situation
\cite{jensen,bhatt1,marley,yaron,balents}.

        In this Letter we report on  the existence of two distinct
states of the FLL, one disordered, the other 
much less disordered (hereafter referred to as the ordered state)
,  with strikingly
large differences between their transport properties. As a result the
system  displays a wide range  of  phenomena such as 
history-dependence, metastability, current-induced annealing and glassy
relaxation. Each of these states is stable
 in its own sharply defined region 
of the (H,T) plane and metastable elsewhere and each  can be
accessed with a  simple  reproducible procedure. 
 Our experiments show directly that the 
metastable state can be annealed into  the  
``equilibrium'' state by applying a current  that depins the FLL. 
  The annealing
kinetics is found to be strongly current dependent, with the annealing time
diverging as the depinning current is approached from above. The variation
of these phenomena with field, temperature,  and
driving current provides direct access to the interplay between static and
dynamic transitions and can elucidate  the role of disorder  in different
parts of the FLL phase diagram. Our results 
can also be used to interpret the rich and complex
history dependence studied earlier in low $T_c$ superconducting
films \cite{kram,kes1}.

    The  history dependence is closely associated with the phenomenon
    of "peak effect".  
    This effect, which is observed in many weak-pinning superconductors,
    is characterized by a sudden
    enhancement of the critical current just below $H_{c2}(T)$ 
    and is  readily observed \cite{bhatt1} as a function
    of H or T. 
    It reflects the
    rapid softening of the elastic moduli of the flux lattice as $H_{c2}(T)$
    is approached, which allows pinning sites to distort the lattice
    more strongly,
    leading to a sharp rise in the critical current.
    This is described in the Larkin-Ovchinnikov(LO) collective pinning
    model \cite{lark} as a reduction in a correlation volume $V_c$ which
    is the characteristic size over which the FLL is ordered.
     In this model the critical current density is given by
     $J_c=B^{-1}(n_p\langle f^2\rangle /V_c)^{\frac{1}{2}}$, where $n_p$
     is the density of pins and f is the elementary pinning interaction.
    Previous studies have shown that as the peak 
region is entered,  the flux dynamics at the onset of
motion changes from elastic flow to plastic flow and finally to fluid flow
as $H_{c2}$ is approached \cite{bhatt1,marley}.

\begin{figure}[btp]
\epsfxsize=3.5in
\epsfbox{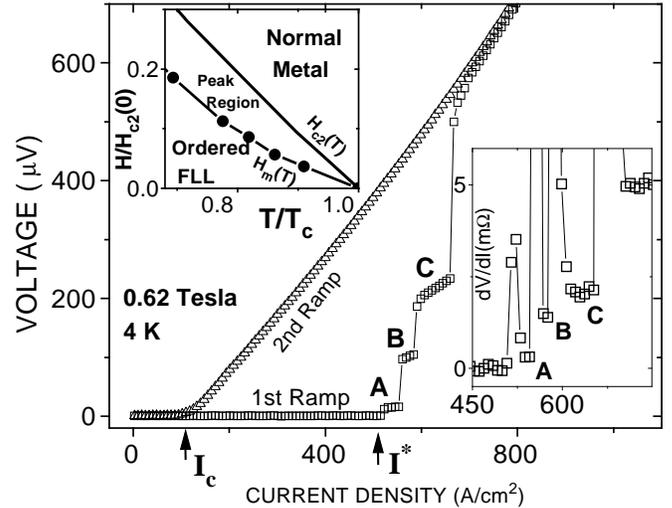}
\protect\caption {I-V curves for the first and second upward ramps of
the applied current after field cooling.     
    Upper inset: The equilibrium phase diagram showing 
    the peak effect region.
    Lower inset: The differential resistance, $dV/dI$ in the 'jumpy' region
near $I^*$. Note the three regions label A,B, and C, which correspond
to the linear segments on the first I-V curve.}
\end{figure}

    Measurements were carried out  on
    two single crystal samples (4x4x0.025  and 1x4x0.025mm) 
    of the layered
    low $T_{c}$ superconductor 2H-$NbSe_{2}$. For this material 
the in-plane penetration depth $\lambda_{\parallel}(0)\sim2000\AA$
    \cite{hnd} and the coherence length $\xi_{\parallel}(0)\sim100\AA$.
    The samples were of
    low purity as indicated by the depressed values of
    $T_{c} = 5.85$ and $ = 6.15K$  ( $T_{c} \sim7.1K$
    in pure 2H-$NbSe_{2}$ ) and low residual
    resistivity ratio $ \sim 9$. Transition widths of 80 and
    50 mK, indicate good homogeneity.
    Resistance measurements
    employed  the standard four probe technique with
   low resistance contacts made with $Ag_{.1}In_{.9}$ solder.
    The current was in the a-b plane and H along the c-axis.

    Fig.1 shows typical I-V curves below the peak region exhibiting 
pronounced history dependence. 
 If the sample is field cooled (FC) through $T_{c}$ with
no applied current, we find a relatively large critical current marked
$I^*$. But this large value is obtained only on the initial ramp up of 
    the current. Once the flux-lines
    are depinned, a much lower threshold current, $I_{c}$ is  found
    as seen on the second ramp up.  Ramping the current down, from
above $I^*$ always gives the low threshold. If, however, 
the sample is zero field cooled (ZFC),  the critical current is low 
even on the initial ramp up (it is actually slightly lower in this
case than in subsequents ramps).
The ratio of $I^*$ to $I_c$ can be  as large as 6, but 
approaches 1 within the peak region (fig. 3). 
At a current above I$^*$, all the I-V curves
join where the slope (dV/dI) agrees with the Bardeen-Stephen \cite{bs}
 value of $ \rho_{n}H/H_{c2} $, indicating nearly free flux flow. 
At even higher currents than shown, deviation in dV/dI occurs
due to sample heating. Heating effects can be large within the 
peak region, but they are readily identifiable and are not
significant in any of the data presented here. 

For a given H and T, the two thresholds, $I_c$ and $I^*$, 
 identify two distinct states 
of the FLL, one more strongly pinned than the other.
The system may be  in one or the other of these states   depending  on 
how the lattice is formed and on its subsequent motion. For example 
FC below the peak region creates a metastable strongly pinned state.  
The system may then be driven into the weakly pinned
state by applying a large current, by changing H, or
by giving the system a mechanical shock. However, changing T 
does not drive the system into the equilibrium state,
except very near $T_c$. 

\begin{figure}[btp]
\epsfxsize=3.5in
\epsfbox{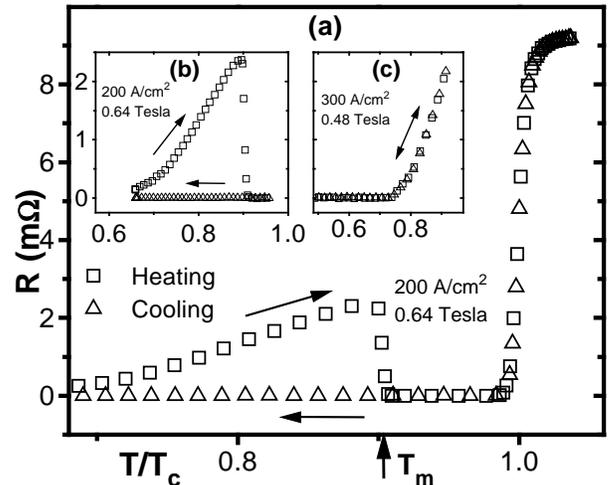}
\protect\caption { Sample resistance for different temperature cyclings.
	 In all cases the lattice was
	 depinned with a large current before the measurement began.
         A much smaller current	 was applied as T was
         cycled. (A.) Cycling T up through $T_{c}$ and then down.
	 	 (B.) Cycling T up halfway into the peak region and
                      then down.
                 (C.) Cycling T down and up without entering peak region.}  
\end{figure}

The peak effect is seen clearly in Fig. 2  
by varying T at  constant  H and fixed current I, $(I_c<I<I^*)$,  
as a sudden drop in resistance 
at a well defined T, marked by an arrow.  
The locus of the resistance jump defines a
transition line, H$_m(T)$ (fig. 1 upper inset)
 which separates the  region where the weakly
pinned state is stable (below the line) from the peak region (above
the line) where the strongly pinned state is stable. 
    The connection between the history dependence and the peak effect
becomes clearer by  cycling the system  halfway into the peak region as
shown in Fig 2b. Here the system is  prepared in the weakly pinned
state by FC below the peak region  and then briefly
applying  a current I$>I^*$. 
    T was then raised while driving a much  lower current
through the sample. 
Upon entering the peak region the voltage drop vanishes, indicating
that flux motion has stopped,  and when the sample is
    cooled again  the FLL remains in the strongly pinned state. 
    This irreversibility is observed
    even if T is  raised into  the peak region only slightly.
    But if  too large a current is used, the behavior
    is reversible.
    \emph{The crucial factor is that the flux motion must stop  in
    the peak region.}
   The same kind of T cycling measurement can be performed without  
entering the peak region, as shown in Fig. 2c.
 The flux motion is 
 stopped by lowering the temperature, but now the response
 varies smoothly and is completely reversible.  

    If the temperature is cycled into the peak region and back 
    with no current flowing, no enhancement of the critical current is
    found unless the temperature is cycled  very near to T$_c$. 
    This implies that the lattice does not enter the strongly pinned
    configuration while in the peak region unless it is driven
    by a current or by thermal fluctuations.
    Thus the weakly pinned state is metastable in the
    peak region except within a narrow fluctuation region
    just below the transition. 
    This metastability is difficult to observe in
    critical current measurements, since
    if a current exceeding the low threshold is applied to this state
    the FLL starts moving, but it quickly finds a more strongly
    pinned state and stops.  
    However, we show below, this transient motion can be
    observed directly in pulse response measurements.

\begin{figure}[btp]
\epsfxsize=3.5in
\epsfbox{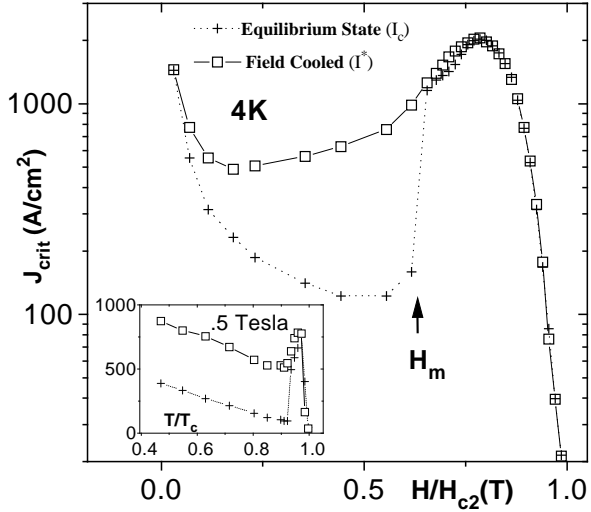}
\protect\caption{Field dependence of the critical current density, $J_{crit}$
         at 4.2K. Inset: Temperature dependence 
         of $J_{crit}$ at .5Tesla.} 
\end{figure}

   We now examine the field dependence of $I_{c}$  and $I^*$ in 
    Fig. 3.  The $I_c$ curve was obtained by an initial 
ZFC followed by I-V measurements for a sequence of fields. 
Obtaining the $I^*$ curve is more difficult, since varying H drives the
system out of the metastable  state, so each point represents 
the threshold current in the 
 first ramp up  after FC through T$_c$. We note that 
in the weakly pinned state, $I_{c}$ has 
    a 1/H dependence below the peak region. In 
 the  LO model 
this behavior reflects the increase in FLL rigidity
 with increasing field and is typical
 of a weakly disordered lattice.  In the FC case, there is a wide
range of fields below the peak over which $I^*$ depends relatively
weakly on field. This indicates that correlations do  not play an
important role and that the FLL must be highly disordered.
 At a field marked $H_m$, $I_c$
shows an abrupt jump up to a value close to the $I^*$ curve. 
At a slightly higher field the two curves merge completely.  A similar
 jump can be seen in the T dependence (fig. 3 inset).

These  results point to  the  origins of the  disordered 
state and of the metastability. The 
 jump in $I_c$ at H$_m$  reflects 
 a sharp decrease in $V_c$ which  
 accompanies a sudden drop or 
vanishing  of an elastic modulus. Thus, as the FLL
enters  the peak region it 
undergoes a  transition to another, less ordered phase \cite{kes2}.
 Field cooling the disordered phase, without sufficient
current to allow  the flux motion  to explore the phase space, 
freezes-in the disordered structure into  a metastable, "supercooled''
 state, leading to the large values of $I^*$. Upon 
driving this metastable   state sufficiently hard, the
system anneals into the equilibrium ordered state which has a large $V_c$
and thus a much smaller value of $I_{crit}$. In the opposite case  when 
 $H_m(T)$ is crossed from the ordered state with pinned
flux-lines,  a metastable ("superheated") ordered structure is 
formed that anneals into a
disordered equilibrium  state when external forces make the flux-lines 
mobile enough to find this state. 

We also note that $I_c(H)$ and $I^*(H)$ join at very low fields (fig. 3).
This is due to the fact that at low fields the FLL becomes dilute,
decreasing the relevance of correlations and enhancing the 
effect of disorder.  What relation, if any,  the merger
bears to a reentrant liquid phase \cite{nelson} remains unclear.  

\begin{figure}[btp]
\epsfxsize=3.5in
\epsfbox{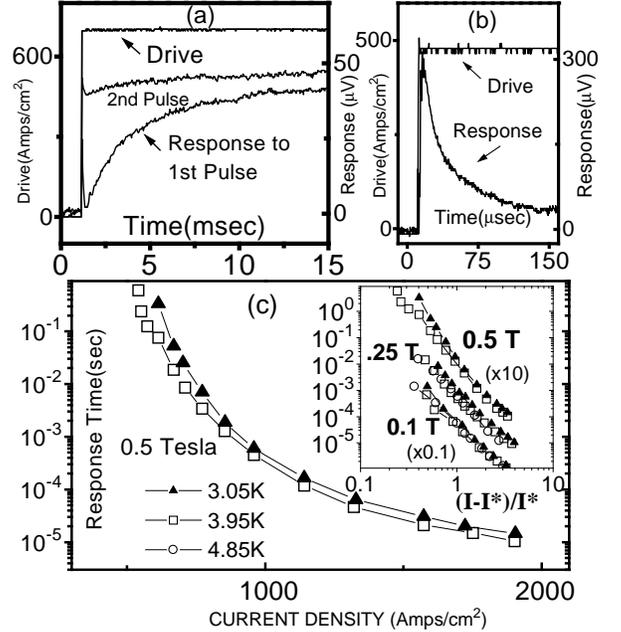}
\protect\caption{Response of the FLL to pulsed currents. (A)  
         Two consecutive pulses after field cooling through $T_{c}$
         to a point below peak region. (The narrow voltage spike is
         due to pickup.)  (B) After warming the weakly pinned 
         state into the peak region. (C) Current dependence of the 
         response time of the strongly pinned state.
         Inset: $t_{r} vs (I-I^*)/I^*$
         for several fields and temperatures. The data for .1 Tesla and
         .5 Tesla were shifted for clarity by the numerical factors shown.}
\end{figure}

      We now demonstrate that the pulsed current technique described
below provides direct access to the annealing
kinetics.  The sample is first prepared in
the disordered   state by  FC  with zero applied current. 
 Then a current pulse $I>I^*$ is 
      applied while the  voltage drop is monitored. 
     Figure 4a shows  the response to  two successive
      pulses: the first exhibits a long response time,  $t_{r}$
       as the disordered  state depins and forms a more  ordered
lattice.\cite{exponent}
      In the second pulse the response is that of  the ordered
      lattice and  is instantaneous. 
      The response time  of the disordered  state exhibits a remarkably
strong  current dependence:   $t_{r}$ changes 
by nearly five orders of magnitude upon changing the pulse height by only a
factor of 3. The data for all  measured H and T  fit a
$t_{r} \propto ((I-I^*)/I^*)^{-\alpha}$, with $ \alpha\sim4$ 
increasing slightly with field
, as shown in Fig. 4c. The data also fit a
$t_{r} \propto exp((I^*/(I-I^*))^\frac{1}{2})$ 
fairly well. The   divergence of the  time scale  suggests the
existence of an underlying dynamical transition at $I^*$.       

We also used this method to observe the complementary situation: 
the metastability of the ordered state in the peak region. 
The system is  prepared in the  ordered state and  heated  into
 the peak region with zero applied current. Subsequently  the current is 
 pulsed with an amplitude I, ($I_c<I<I^*$). We find an initial 
response which is instantaneous, as in  the ordered phase, but
drops   quickly and dramatically, as the moving lattice finds a more
strongly pinned configuration (fig4b). This indicates that both 
superheating and supercooling occur across the equilibrium phase
boundary at $H_m(T)$. 

    The results presented here provide a new basis for understanding 
    the depinning of disordered flux lattices.  A  commonly used model
    of depinning is as a process of driving the FLL
    over a static barrier, whose height is taken to be   the DC critical
    driving force. However, the net force on a FLL gives its instantaneous
    center of mass velocity and thus, according to the model, 
    if  the system is driven with a current 
    $I>I^*$   the FLL should  respond instantaneously. The 
    existence of long response times  
    shows that the pinning force can momentarily be much larger than
    the value implied by the size of $I^*$ and that the depinning transition 
    is more complicated.
    This is not surprising since the model neglects the internal
    degrees of freedom of the lattice, which are important here.
    A more accurate description of depinning for the 
    disordered lattice is as a dynamical  transition
    with long relaxation times associated with reordering. 

    The  data in the 'jumpy' transition region just above
    $I^*$ (fig. 1) in 
    the I-V curve for the disordered state indicates that
 the onset of motion is in the form of channels
    of flowing flux-lines \cite{jensen,bhatt1}. In this scenario, the
jumps are the result of the sudden depinning of some fraction of the
flux-lines, while the linear segments between jumps reflect the
increasing speed of the moving flux-lines as the current is raised. 
Estimates of the moving fraction, f can be made by comparing the
voltage and differential resistance, $R_d$ (Fig. 1 lower inset) in the
linear segments to their 
values in the equilibrium state. For the three segments shown in
Fig. 1, we obtain f=.04,.22,.43 and f=.06,.30,.39 from the values of
V and $R_d$ respectively.

The existence of motion in channels may provide 
insight into the long response times.
One possibility is that when a current $I > I^*$ is applied,
 a few flux lines start moving along 
continuous channels which open across the
sample, with the number of open channels depending on I.
 The moving segments of the  FLL 
order and expand outwards  until they engulf the entire sample 
at which point the system is indistinguishable from the moving weakly
pinned phase.  This growth process could lead to the long
relaxation times.
 The apparent
    divergence of t$_r$ at $I^*$ may reflect a divergence in
    the length scale of the separation between adjacent channels.

    In conclusion, we have observed that the FLL can be prepared in 
 metastable states which are sufficiently robust to 
 allow detailed study of their   transport
properties.   Annealing of the metastable  states
    reveals  glassy relaxation times which are strongly dependent
    on the driving current.   While 
these results were obtained   in 2H-$NbSe_2$, they appear to reflect 
only  the properties of the FLL and the disorder,  and therefore 
should be valid  in other superconductors with comparable levels of
disorder. We hope these results  will stimulate 
 experimental and theoretical work on 
the  annealing mechanism  and how it is
affected by disorder.

 We acknowledge discussions with L. Balents, P. Chaikin, J. Clem,
M.P.A. Fisher, L. Ioffe, X.S. Ling, A. Schofield, C. Tang, and
V. Vinokur.
We thank P. Kes for bringing useful references to our attention.


\begin{references}
\bibitem{exponent} For I not too close to I$^*$ the voltage
evolves towards the equilibrium value of the 
ordered phase $V_f$ with a simple  time dependence
$V=V_f(1-e^{-t/\tau_r})$. For I  only slightly above I$^*$ 
 the response is  characterized by  more than one time scale.
\end{references}
\end{document}